\documentstyle[PASJadd,psfig]{PASJ95}
\draft
\newcommand{\vol}{}
\newcommand{\no}{}
\newcommand{\lett}{}
\newcommand{\spage}{}
\newcommand{\rdate}{1997 June 20}
\newcommand{\adate}{1997 December 2}

\begin{document}

\title{Initial Shock Waves for Explosive Nucleosynthesis in Type II Supernova}

\author{Shigehiro {\sc Nagataki},$^1$ Masa-aki {\sc Hashimoto},$^2$,
and Shoichi {\sc Yamada}$^{1,3}$
\\[12pt]
$^1$ {\it Department of Physics, School of Science, the University
of Tokyo, 7-3-1 Hongo, Bunkyoku, Tokyo 113 }\\
{\it E-mail(TY): Nagataki@utaphp1.phys.s.u-tokyo.ac.jp}\\
$^2$ {\it Department of Physics, Faculty of Science,
Kyusyu University, Ropponmatsu, Fukuoka 810} \\
$^3$ {\it 
Max-Planck-Institute f\"ur Physik und Astrophysik,
Karl-Schwarzschild Strasse 1, D-8046, Garching bei M\"unchen, Germany 
}
}

\abst{
We have performed 1-dimensional calculations for explosive nucleosynthesis 
in collapse-driven supernova
%%%%%%%%%%%%%%%%%%%%%%%%%%%
and investigated its sensitivity to the initial form of the shock wave.
We have found the tendency that the peak temperature becomes higher around the 
mass cut if the input energy is injected more in the form of
kinetic energy rather than internal energy. Then, the mass cut becomes
larger, and, as a result, neutron-rich matter is less included in the
ejecta; this is
favorable for producing the observational data compared with a
previous model. Our results imply that the standard method to treat various 
processes for stellar evolution, such as convection and electron capture
during the silicon burning stage, are still compatible with the
calculation of explosive nucleosynthesis.
%%%%%%%%%%%%%%%%%%%%%%%%%%%%%%%%%%%%%%%%%%%%%%%%%%%%
}

\kword{
Nucleosynthesis --- Supernovae: general --- Supernovae:
individual (SN 1987A) }

\maketitle
\thispagestyle{headings}

%%%%%%%%%%%%%%%%%%%%%%%%%%%%%%%%
\section{Introduction}
%%%%%%%%%%%%%%%%%%%%%%%%%%%%%%%%
\setcounter{equation}{0}

\indent

Elements heavier than $\rm ^{12} C$ are mainly synthesized
during the hydrostatic evolution of stars and supernova explosions.
It can be said that massive stars play an important role concerning
the chemical
evolution of the Galaxy, because they produce most elements of $Z < 30$. 
%%%%%%%%%%%%%%%%%%%%%
Until today, many calculations about stellar evolution and
supernova explosion have been performed, and 
the compositions in the ejecta have been predicted. However, 
consistent calculations from stellar evolution to explosions still do not
exist (e.g., Arnett 1996).
In this paper we pay attention to collapse-driven supernova,
which is regarded as the death of a massive star whose main sequence
mass exceeds 8-times the solar mass ($M_{\odot}$) (e.g., Hashimoto 1995).

Because of uncertainty concerning the collapse-driven supernova 
mechanism, calculations of stellar evolution and explosive
nucleosynthesis during a supernova explosion have been done
separately; a shock wave is artificially generated at the inner region of 
the core of the progenitor (Hashimoto et al. 1989).

%%%%%%%%%%%%%%%%%%%%%%%%%%%%%%%%%%%%%%%%%%%%%%%%%%%
Historically speaking, there have been two ways to generate a shock wave,
as analyzed in detail by Aufderheide et al. (1991). One is called an
internal-energy bomb (hereafter referred to as the bomb); the other is
called a piston. In the approach of the bomb, the 
input energy is deposited in the form of
internal energy at the inner-most edge
 of the calculation region,
so that all of the energy deposited propagates outward. 
On the other hand, in the piston method, the inner-most edge 
is moved as a 
%%%%%%%%%%%%%%%%%%%%%%%%%%
piston, so that the final explosion energy amounts to about $10^{51}$ erg
(Woosley, Weaver 1995).

Though these methods may be a good approximation, and explain many
observational data, they have difficulty concerning the peak
temperature during the early
phase of the shock. Aufderheide et al. (1991) studied 
both the bomb and the piston methods to determine the influence on the
initiation
and propagation of the shock wave. They also investigated the effect due to the
different launching time of the shock wave: at $t = 0$ (the initial 
presupernova model: uncollapsed model) 
and after 0.28 s of core collapse (collapsed model). 
They found that up to a $10 \%$ 
difference 
in the major abundances from
the 
different shock initiation schemes 
(the bomb or the piston) 
and up to $30 \%$
due to a
variation in the launching time. 
The main result of their investigation is that the peak temperatures
are different between the two methods in the early
phase of shock propagation, although they converge as the shock
waves propagate forward. 
This is because too much energy is deposited in the internal energy for the
bomb and too much in the kinetic energy for the piston. 
As a result, there is uncertainty concerning the chemical composition
in the inner
region of the ejecta of $M < 2 M_{\odot}$, whose progenitor mass is assumed to
be a 20 $M_{\odot}$ star. 

%%%%%%%%%%%%%%%%%%%%%%%%
We now make an important comment concerning the presupernova model.
Aufderheide et al. used the 6 $M_{\odot}$ helium-core model of Nomoto and
Hashimoto (1988), which may have a problem of excessive neutronization. 
During the silicon burning, neutronization occurs below the O-rich
layer due to electron capture. 
In this model some products of explosive nucleosynthesis become
too neutron-rich. For example, the ratio of the mass fraction $\rm
^{58}Ni/^{56}Ni$ becomes much higher than the observation of SN 1987A
(e.g., Hashimoto 1995). 

%%%%%%%%%%%%%%%%%%%%
Considering the large uncertainty of convective theory, Hashimoto
(1995), including Aufderheide et al., have made the value
of $Y_{\rm e}$  higher artificially in the Si-rich layer, so as to reproduce 
the observational ratios of the mass fraction for important nuclei,
such as $\rm
^{57}Ni/^{56}Ni$ and $\rm ^{58}Ni/^{56}Ni$.
This means that the convection and/or the electron capture are limitted
in the Si-rich layer, and that the treatments of
the convection and the electron capture during stellar evolution are
inconsistent with the calculations of explosive nucleosynthesis
because the discontinuity of $Y_{\rm e}$ in the progenitor is formed at the
outer boundary of the convective shell (Thielemann et al. 1990;
Hashimoto 1995). For example, the value of $Y_{\rm e}$ for $M > 1.607
M_{\odot}$ (=0.494) is artificially 
changed to that of $M > 1.637 M_{\odot}$ (=0.499) in Hashimoto (1995).
We also note that the Schwartzschild criterion is adopted in Nomoto \& 
Hashimoto (1988) for convective stability, neglecting both of overshooting
and semiconvection, which means that the convective motion is fairly
suppressed in the calculation of the progenitor.

In this paper, 
reversing the argument, 
we make use of the fact that the product of the
explosive nucleosynthesis has uncertainty 
due to the poorly known initial form of the shock wave,
and investigate whether we can 
find 
an
initial shock 
wave which reproduces 
the observational data with the presupernova model unchanged. 
In particular, Aufderheide et al. (1991)
studied only the bomb or the piston, and did not examine their
combination. 
In this paper we show that an initial shock wave 
comprising a proper combination of kinetic and internal energy can reproduce
the observed chemical abundances. 
%%%%%%%%%%%%%%%%%%%%%%%%%%%%%%%%%%%%%%%%%%%%%%%%%%%%%%
This means that the electron-capture rates of Fuller et al. (1980, 1982) and
convection
using the Schwarzschild criterion during stellar evolution are
still compatible with the observational data of the explosive nucleosynthesis
products. Moreover, 
the form of the initial shock wave found in this study should be the
outcome of core-collapse calculations.

In section 2 we explain explosive nucleosynthesis in
a 
collapse-driven supernova explosion.
In section 3, the methods adopted to calculate the explosive
nucleosynthesis
are 
described.
We show the results in section 4.
A summary is 
presented 
in section 5.

%%%%%%%%%%%%%%%%%%%%%%%%%%%%%%%%%%%%%%%%%%%%%%%%%%%%%%%%%%%
\section{Explosive Nucleosynthesis} \label{explosive}
%%%%%%%%%%%%%%%%%%%%%%%%%%%%%%%%%%%%%%%%%%%%%%%%%%%%%%%%%%%

%-------------------------------------------
\subsection{General Feature} \label{general}
%-------------------------------------------

\indent

In this section we give the general features of explosive
nucleosynthesis that are common to collapse-driven supernova phenomena.
The temperature and entropy 
per nucleon become high after the passage of a shock
wave. The increase in temperature $T$ causes various nuclear reactions
that have been blocked by the Coulomb barrier in the nuclear
burning stage during the hydrostatic stellar evolution.
On the other hand, heavy elements disintegrate due to 
the high entropy 
per nucleon.
This leads to the production of light nuclei, such as
n, p, and $^4 \rm He$. Usually, the final chemical
composition is not in chemical equilibrium, and depends on the time
variation of $\rho$ and $T$ after passage of the shock wave.

%%%%%%%%%%%%%%%%%%%%%
Here, we focus on the death of a star having $\sim 20
M_{\odot}$ in the main-sequence stage, like SN 1987A. In this case, the
mass cut is assumed to be located in the Si-rich layer of the progenitor. 
Hence, we explain the explosive nucleosynthesis in the Si-rich
layer and O-rich layer, where explosive nucleosynthesis in our interest
occurs.
%%%%%%%%%%%%%%%%%%%%%%%

(i) Explosive Si-burning  \\

\indent
In the  explosive Si-burning layer of $T \approx 5\times10^9 \rm K$,
an alpha-rich freezeout occurs. After passage of the shock wave,
most nuclei are at first photo-disintegrated. Then, nuclei begin to
recombine 
with each other as the temperature falls along with expansion.
At such a high temperature, all of the Coulomb
barriers can be overcome, and Fe-group nuclei are mainly synthesized. 
%%%%%%%%%%%%%%%%%%%%%%%%%%%%%%%%%%%%%%%%%%%%%%%%%%%%%%%%
It should be noted that how the produced elements are neutron rich is very
sensitive to the distribution of $Y_{\rm e}$ in the Si-rich layer, which is formed
during the silicon shell-burning stage. 
For an electron fraction of 
$Y_{e}\; \ge \;0.493$ the most abundant nucleus is the doubly-magic nucleus
$^{56}\rm Ni$, which has the largest binding energy per nucleon for $N
=Z$. On the other hand, if $Y_{\rm e} < 0.493$, the most abundant 
nucleus becomes $\rm ^{58}Ni$.
Important radioactive nuclei, $\rm ^{57}Ni$ and $\rm ^{44}Ti$, which are 
 sources of heating the ejecta, are also synthesized in this layer.
%%%%%%%%%%%%%%%%%%%%%%%%%%%%%%%%%%%%%%%%%%%%%%%%%%%%%%%%%%%%%

(ii) Explosive O-burning  \\

\indent
At the inner-most region of the O-rich layer, the peak temperature
becomes sufficiently high to produce Fe-group nuclei. The important thing to 
be emphasized is that $Y_{\rm e}$ is nearly 0.5 in this layer, because electron
capture does not work effectively during the oxygen shell burning stage.
As a result, it is not $^{58}\rm Ni$, but $^{56}\rm Ni$, that is
produced mainly at the inner most region of the O-rich layer.

As the shock wave decays, Fe-group nuclei cannot be produced through
explosive burning because of the bottle neck at the proton magic number
$Z=20$. A temperature larger than $\sim\;3.3\times10^9\;\rm K$
leads to a quasi-statistical equilibrium (QSE) among nuclei in the range $ 28\;
<\;A\;<\;45$ in mass number.

%----------------------------------------------------
\subsection{Observational Data} \label{observation}
%----------------------------------------------------

\indent

It is only SN 1987A in the Large Magellanic Cloud that has provided the
most precise data to prove the validity of 
explosive nucleosynthesis calculations.
For example, the mass of $^{56} \rm Ni$ has been estimated to be $0.07$--$0.076
M _{\odot}$ on the basis of a luminosity study (Shigeyama et al. 1988;
Woosley, Weaver 1988).
The values of $\langle \rm ^{57}Ni/^{56}Ni \rangle$ and $ \langle \rm
^{58}Ni/^{56}Ni \rangle$ are also determined from the observation.
These values are defined as below:
\begin{eqnarray*}
 \langle\rm ^{57}Ni/^{56}Ni\rangle \equiv[\it X(\rm ^{57}Ni)/ \it X(\rm
^{56}Ni)]/[\it X(\rm ^{57}Fe)/\it X(\rm ^{56}Fe)]_{\odot},  \\
 \langle\rm ^{58}Ni/^{56}Ni\rangle \equiv[\it X(\rm ^{58}Ni)/ \it X(\rm
^{56}Ni)]/[\it X(\rm ^{58}Ni)/\it X(\rm ^{56}Fe)]_{\odot},  \\
\end{eqnarray*}
where $X$ denotes the mass fraction.  
The ratio of $^{57} \rm Ni$ to $^{56} \rm Ni$ has been determined from the
X-ray light curve to be $1.5 \pm 0.5$ times the solar $\rm
^{57}Fe$/$\rm ^{56}Fe$ ratio (Kurfess et al. 1992).
From the spectroscopic
observation of SN 1987A, the mass of $\rm ^{58}Ni$ in SN 1987A has
been estimated to be $\sim 0.0022$--$0.003 M_{\odot}$ (Rank et al. 1988;
Witterborn et al. 1989; Aitken et al. 1988; Meikle et al. 1989;
Danziger et al. 1991),
which gives the ratio $ \langle\rm ^{58}Ni/^{56}Ni\rangle $
$\sim$ 0.7--1.0.
In section 4, we will present the initial conditions of the shock wave 
that satisfy these three observational constraints.

%%%%%%%%%%%%%%%%%%%%%%%%%%%%%%%%%%%%%%%%%%%%%%%%%%%%%%
\section{Models and Calculations} \label{calculation}
%%%%%%%%%%%%%%%%%%%%%%%%%%%%%%%%%%%%%%%%%%%%%%%%%%%%%%

%-----------------------------------
\subsection{Hydrodynamics and Nuclear Reaction Network} \label{hydro}
%-----------------------------------

\indent

In this section we explain our numerical calculation method.
We performed 1-dimensional hydrodynamical calculations.
The number of meshes is 300 in the radial direction.
The inner and outer most radius are set to be $10^{8} \ \rm cm$ and $2
\times 10^{10} \ \rm cm$, respectively.
We adopt the Roe method for the calculation (Roe 1981; Yamada, Sato 1994).
We use the equation of state;
\begin{eqnarray}
P = \frac{1}{3} a T^{4} + \frac{\rho k_{\rm B} T}{ A_{\mu} m_u },
\end{eqnarray}
where $a$, $ k_{\rm B}$, $ A_{\mu}$, and $ m_u$
are the radiation constant, the Boltzmann
constant, the mean atomic weight, and the atomic mass unit,
respectively. 

We also use a test-particle method to see the variations of 
($\rho , T$) in the Lagrangian coordinate. 
%%%%%%%%%%%%%%%%%%%%%%%%%%%%%%%
Let us  explain this method.
It is assumed that test particles are at rest in the beginning and move with
the local velocity at their positions after the passage of a shock
wave. We can thus calculate each particle's path by integrating 
$\displaystyle{\frac{d \vec{x}}{d t}} 
= \vec{v}(t, \vec{x}) $, where
the local velocity $\vec{v}(t, \vec{x})$ is given from the
hydrodynamical calculations mentioned above. The density and
temperature of a test particle at each time step are obtained through an
interpolation between the Eulerian meshes where the particle is found at
the moment. 
We can thus obtain information about ($\rho , T$) for the Lagrangian
coordinate, while
preserving the time variation of the
density and temperature along each trajectory of the test particles.

\begin{figure}
\begin{center}
   \leavevmode\psfig{figure=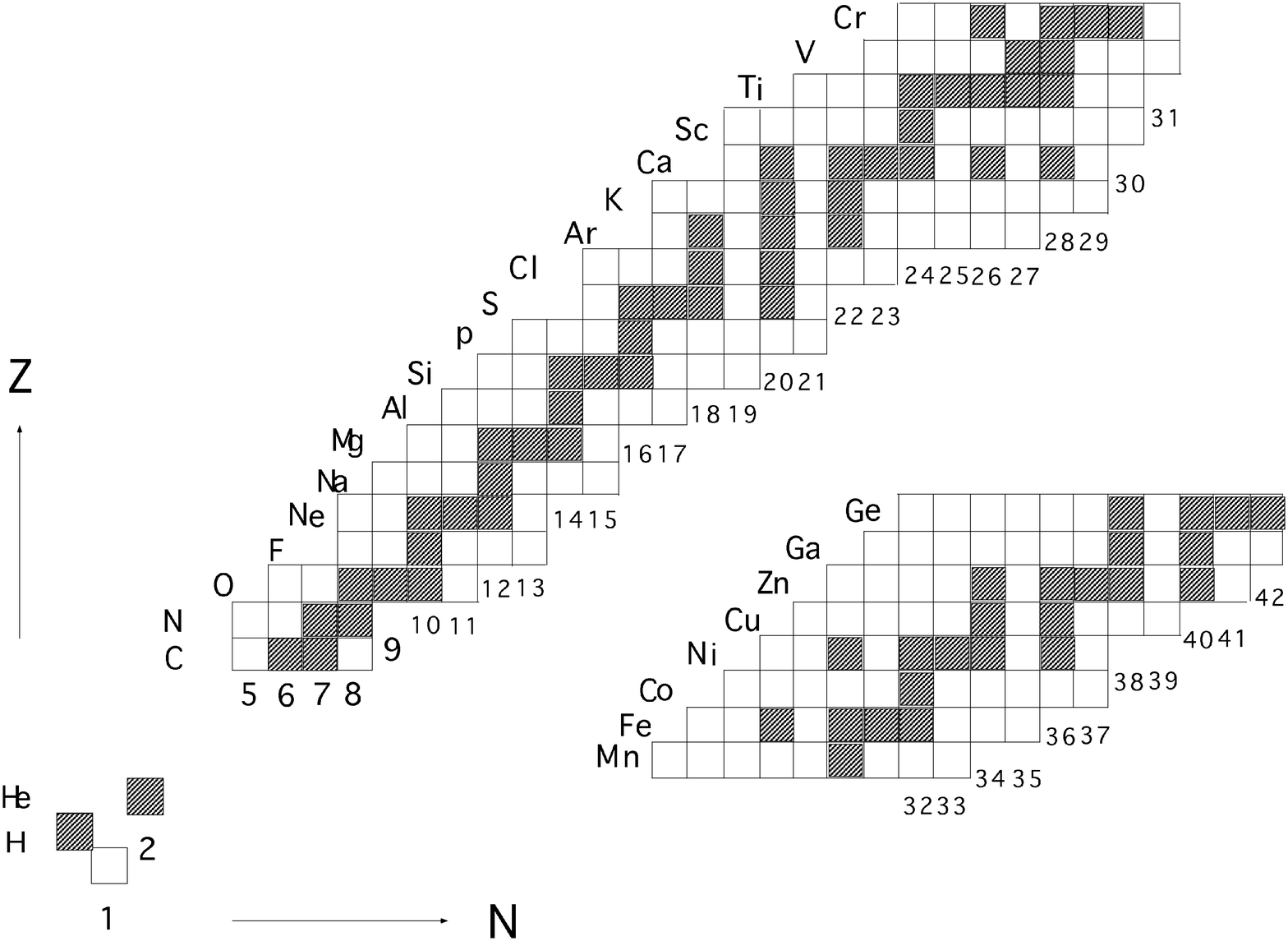,height=14cm,angle=0 }
\end{center}
\caption{}
\label{table}
\end{figure}

Next, we calculate the explosive nucleosynthesis using the time-dependent data
of $(\rho,T)$. Since the system is not in
chemical equilibrium, we must calculate the change in the chemical
composition using a nuclear reaction network containing 
242 nuclear species (see figure~\ref{table}, Hashimoto et al. 1989). 
%%%%%%%%%%%%%%%%%%%%%

It is noted that we assume that the system is adiabatic after the passage of
a shock wave,
because the entropy produced during explosive nucleosynthesis is
much smaller than that generated by the shock wave. 
Thus, nucleosynthesis calculations are carried out separately, so-called
$post\ processing$, for each trajectory of the test particles using the  
nuclear reaction network.
Refer to Nagataki et al. (1997) for details concerning the numerical
method, which has also
been used in a calculation of asymmetric explosive nucleosynthesis.
%%%%%%%%%%%%%%%%%%%%%%%%%%%%%%%%%%%%%%%%%%%%%%%%%%

%-----------------------------------
\subsection{Initial Condition} \label{init}
%-----------------------------------

\indent

The progenitor of SN 1987A, Sk $-$69$^{\circ}$202, is thought to
have had the mass of $\sim 20 M_{\odot}$ in the main-sequence stage
(Shigeyama et al. 1988; Woosley, Weaver 1988) and had an $\sim$
(6$\pm$1)$M_{\odot}$ helium core (Woosley 1988).
%%%%%%%%%%%%%%%%%%%%%%%%%%%%
We used the presupernova model just before the collapse obtained 
from the evolution of a helium core of 6 $M_{\odot}$ (Nomoto,
Hashimoto 1988) as the initial model for density and compositions.
We stress that calculations were performed while keeping the original
distribution
of $Y_{\rm e}$.
Table 1 shows the radii of the Fe/Si, Si/O, and O/He
interfaces in this model, where a discontinuity of the compositions exists
due to convective shell burning.

We now explain the way that the shock wave is initiated in this study.
We deposit some amount of energy (input energy) as a combination of the
internal energy and kinetic energy at the inner-most edge of the
calculation region. 
Both the internal energy and 
the initial velocity are assumed to be 
proportional to the radius. Although the way of injecting the
explosion energy is
artifitial, we note that the velocity of the Sedov solution in uniform
density is proportional to the radius (Sedov 1959).
The initial velocity is also
assumed to be radial. The input energy is set to be $2.0 \times
10^{51} \rm erg$ (2 foes), which is appropriate for SN 1987A, since the final
explosion energy, 1 foe, is the sum of the input energy and the
gravitational binding energy
above the mass cut. 
The ratio of the internal energy to the kinetic energy is
changed parametrically to see its effect on the nucleosynthesis. 
We also studied the effect of a change in the range of
the region where the input energy is injected. 
%%%%%%%%%%%%%%%%%%%%%%%%%%%%%%%%%%%%%%%%%%%%%%%%
We note that locating the input energy further away from the center
implies setting the final explosion energy larger.
The initial conditions explored in this study are summarised in table 2.

%%%%%%%%%%%%%%%%%%%%%%%%%%%%%%%%
\section{Results}
%%%%%%%%%%%%%%%%%%%%%%%%%%%%%%%%

\indent

\begin{figure}
\begin{center}
   \leavevmode\psfig{figure=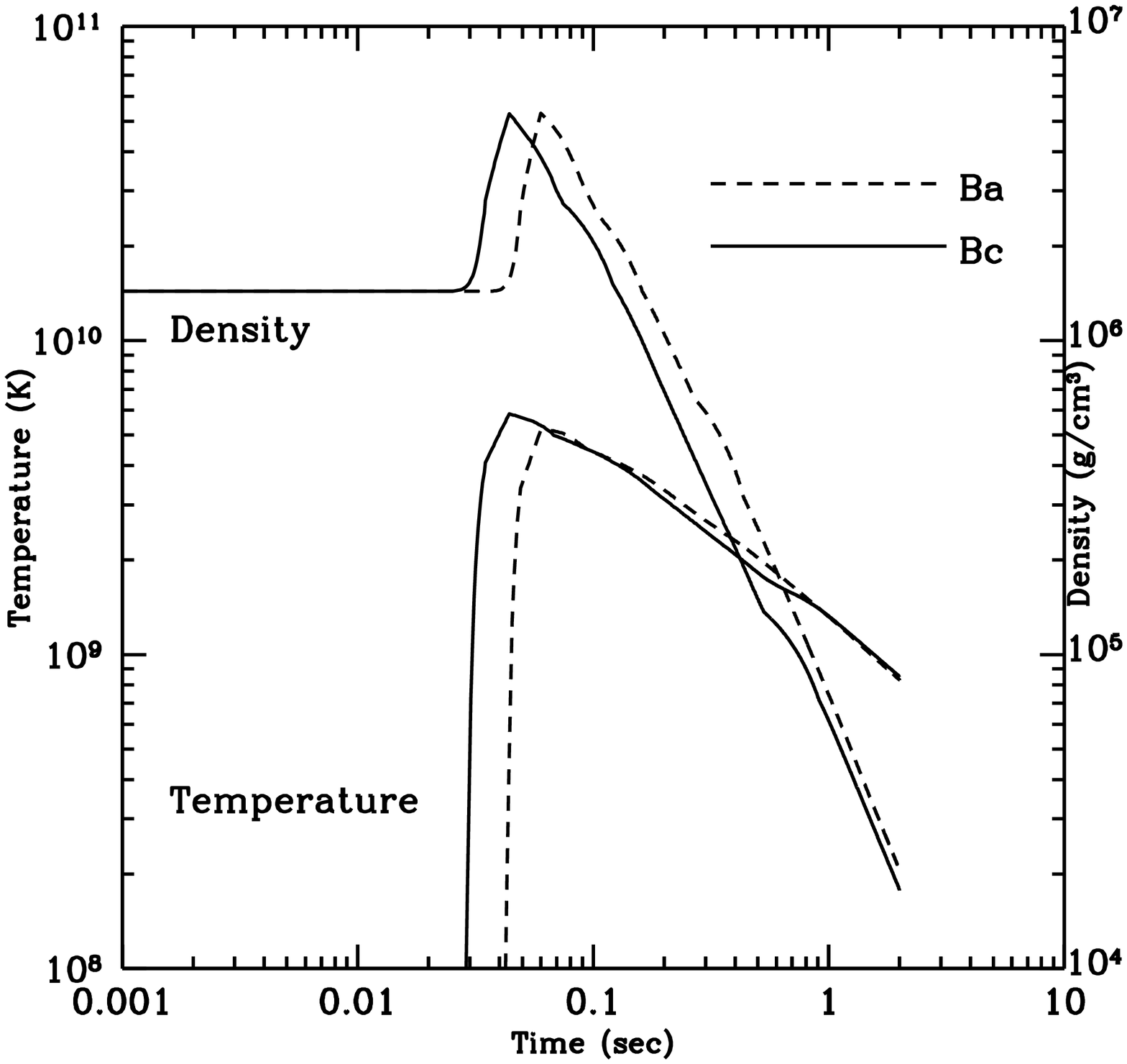,height=18cm,angle=0 }
\end{center}
\caption{}
\label{rhotemp}
\end{figure}

\begin{figure}
\begin{center}
   \leavevmode\psfig{figure=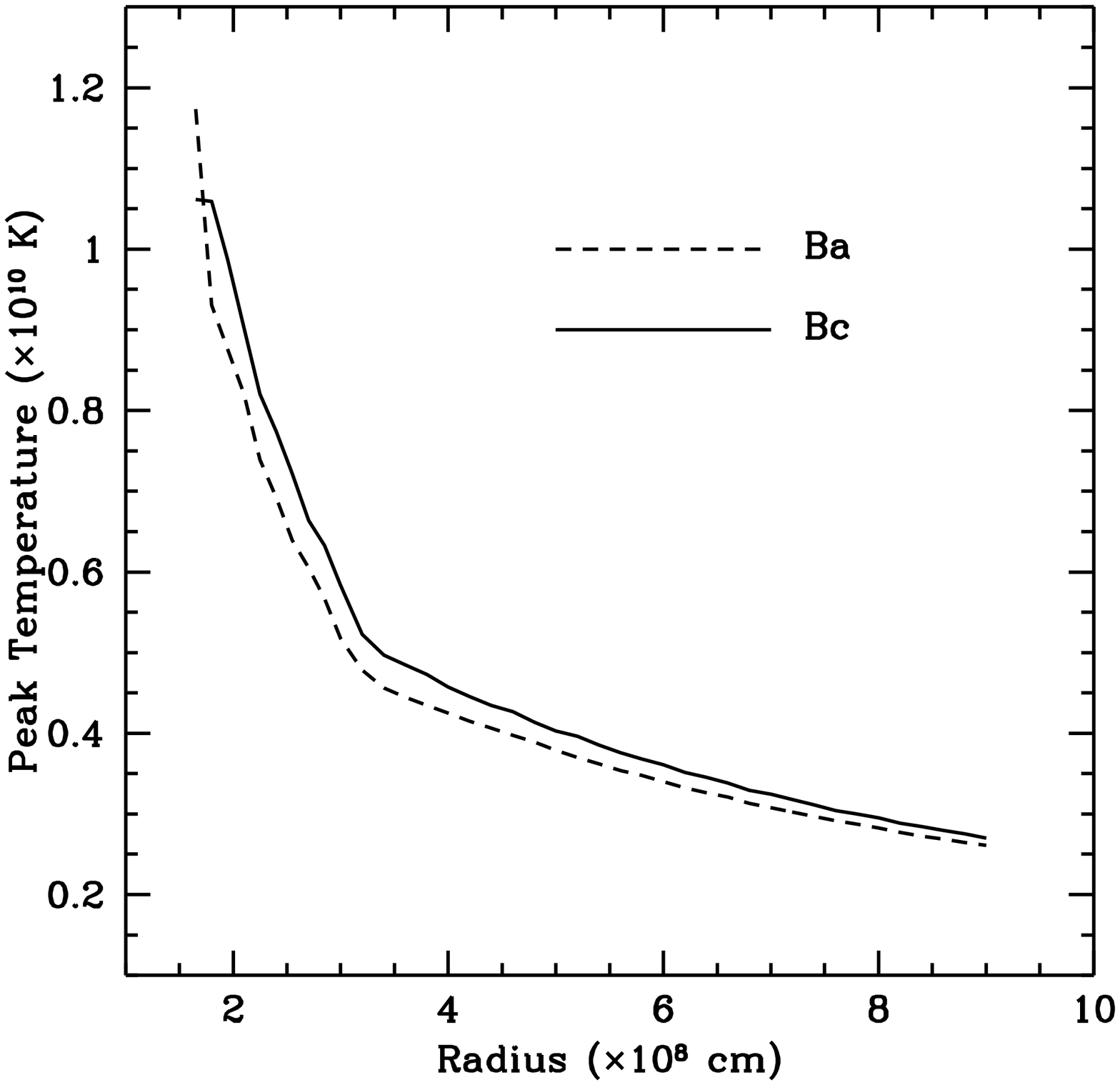,height=18cm,angle=0 }
\end{center}
\caption{}
\label{peaktemp}
\end{figure}

%%%%%%%%%%%%%%%%%%%%%%
As stated in section 3, the initial shock waves of models Cs (Ca--Cc) are
strongest and those of As (Aa--Ac) are weakest. In this section we consider
the results of the intermediate strength of the shock waves, that is,
the results of Bs (Ba--Bc) for representation. In particular, we
discuss models 
Ba and Bc in order to understand the effect of the ratio of the initial kinetic energy to the
initial thermal energy.
The products from explosive nucleosynthesis depend crucially on the peak
temperature, which is affected by the method of shock initiation.
To see the difference of $(\rho,T)$ during shock-wave propagation, we show
in figure~\ref{rhotemp} changes in $(\rho,T)$ with time for a test particle
initially placed at the boundary between the O-rich and
Si-rich layer ($r \; = \; 3.0\times 10^8$ cm) for models Ba and Bc. 
We can see that the peak
temperature is higher in 
model Bc at $t \sim 0.04$ s than that in model Ba at $t \sim 0.06$ s. 
To see this tendency more clearly, we show the peak temperature for 
models Ba and Bc 
in figure~\ref{peaktemp}. The abscissa 
represents 
the initial 
position of each test particle. We can see that the peak temperature
is higher at first ($r < 1.8 \times 10^8$ cm) in 
model Ba.
On the other hand, the peak temperature of model Bc is higher in the rest 
of the region.
We can give an explanation to this phenomenon 
as follows.
The temperature is 
simply determined by  
the internal-energy density 
behind the
shock wave, because this region is photon-dominated.
Since the input energy is deposited 
only in the form of internal energy, the peak temperature becomes 
higher at first in model Ba.
However, as the shock wave proceeds forward, the shock becomes weaker
compared that in model Bc, since more energy is dissipated at smaller radii.
As a result, the peak temperature of model Ba becomes lower
for $r > 1.8 \times 10^8$ cm compared with model Bc.
This tendency is consistent with the result of 
Aufderheide et al. (1991) as can be seen from their figure 7.

We now examine the effect on  explosive nucleosynthesis.
We calculated the total amount of heavy elements in the range
$A = 16$--$73$ and compared them with 
the solar system abundances.
Some comments are necessary concerning this analysis.
First, all unstable
nuclei produced were assumed to decay to
stable ones when compared with the solar values.
Second, the mass cut was determined so as to contain $0.07 M_{\odot}$ of
$\rm ^{56} Ni$ in the ejecta.
The mass cut for each model is shown in table 3 along with the peak 
temperature at the Si/O interface.
We also note that the initial mass function (IMF), the 
chemical composition in the ejecta for each mass of the progenitors,
and the ratio of Type I to Type II supernova are necessary when a
comparison with the solar-system abundances is made in relation to the
chemical evolution of the galaxy. 
In the present investigation, we can only see to what extent the
abundances could be affected by the form of the initial shock wave.

\begin{figure}
\begin{center}
   \leavevmode\psfig{figure=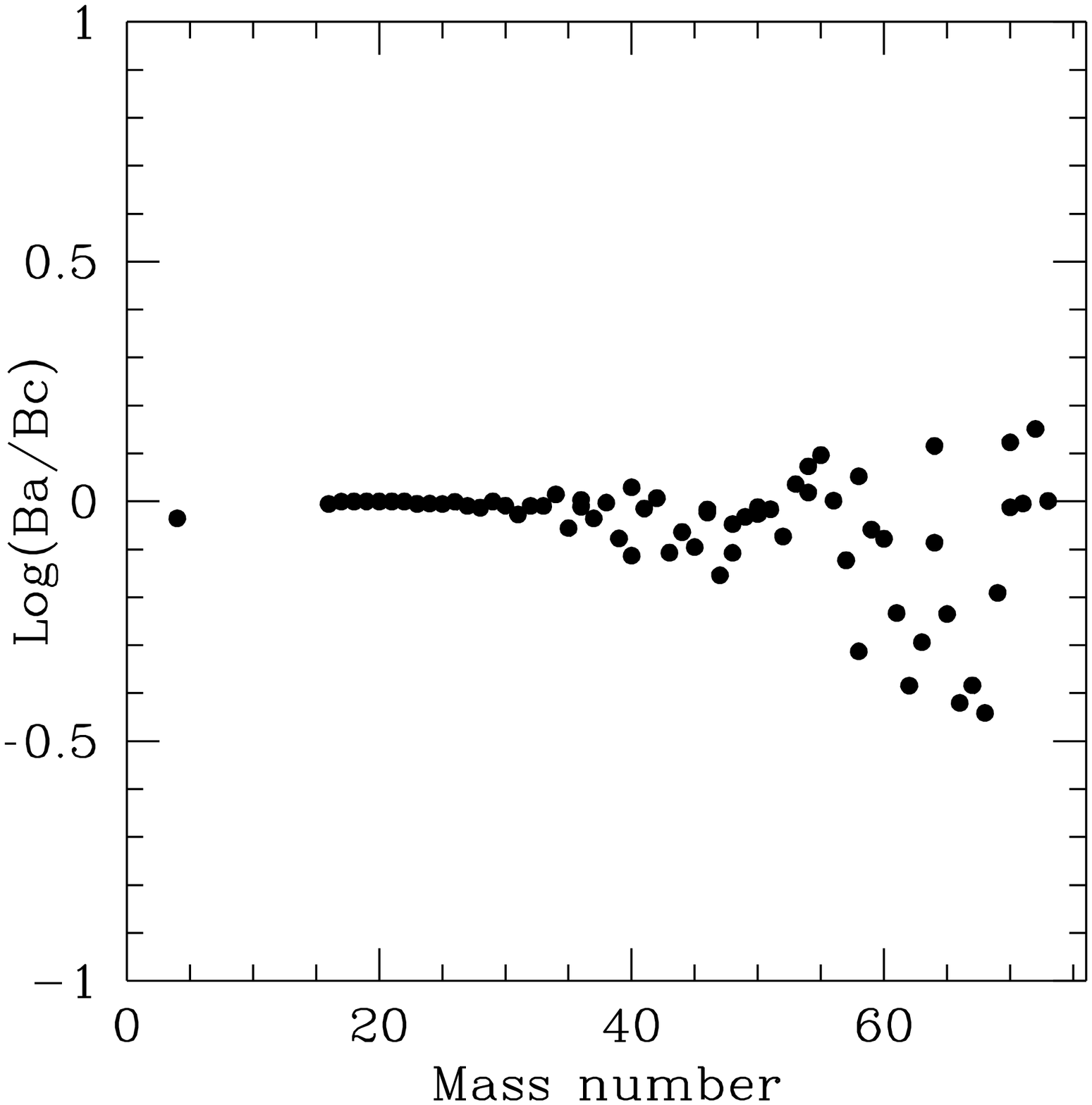,height=18cm,angle=0 }
\end{center}
\caption{}
\label{warubabc}
\end{figure}

\begin{figure}
\begin{center}
   \leavevmode\psfig{figure=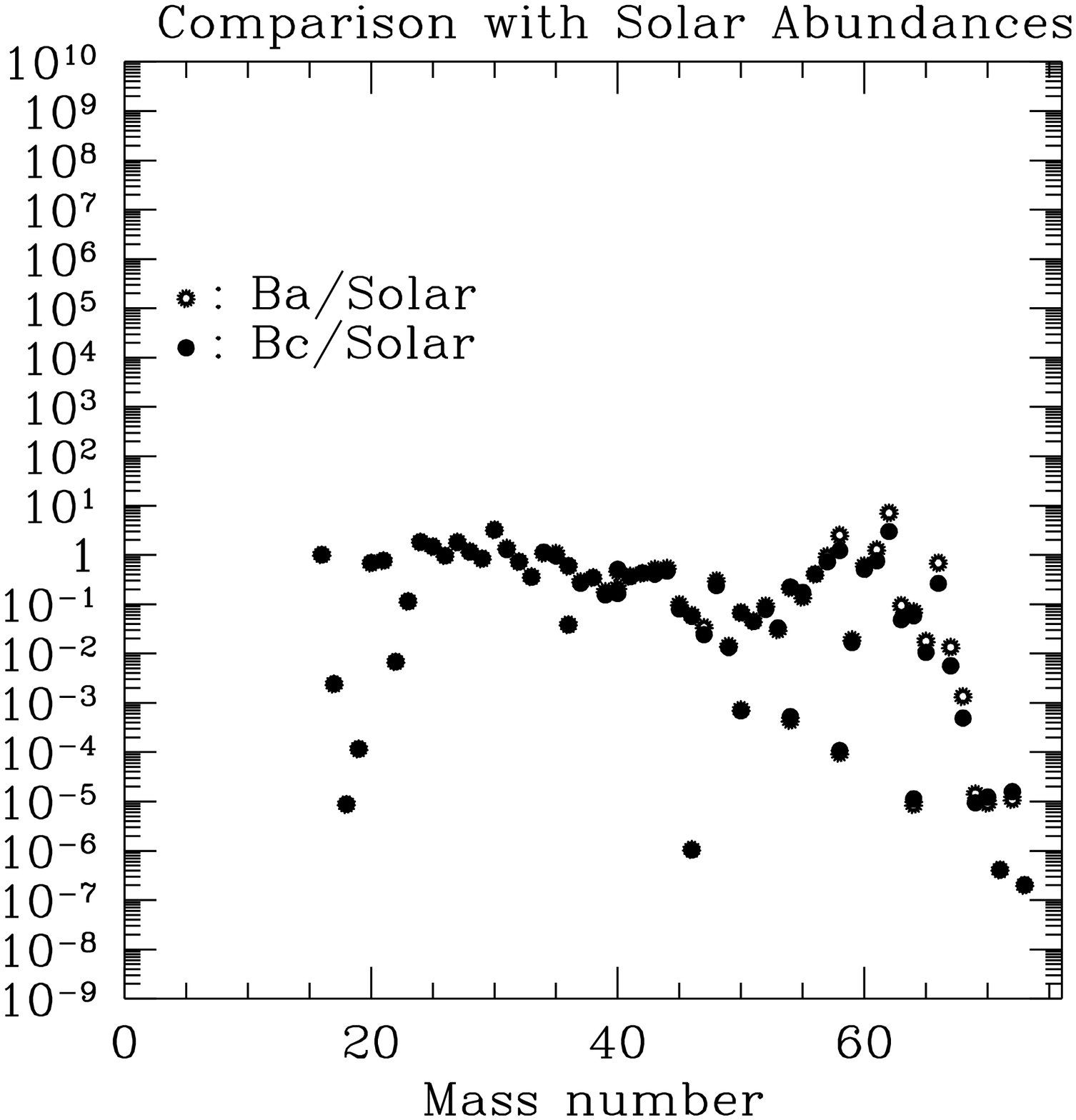,height=18cm,angle=0 }
\end{center}
\caption{}
\label{solar}
\end{figure}

Figures ~\ref{warubabc} and ~\ref{solar} show the results
for 
models Ba and Bc. 
The former 
shows a comparison of the composition for $A=16$--$73$
between
models Ba and Bc.
The dots denote the mass fraction ratio of model Bc to Ba.
The latter  
illustrates a comparison of the abundances of
ejected nuclei with the solar values (normalized at $\rm ^{16} O$).
The open circles 
represent 
Ba/Solar and the dots 
correspond to 
Bc/Solar, respectively.
It is evident from
figure \ref{warubabc}
that the number of nuclei in the range $A=16$--$50$ is almost the same
between the two models and within a factor of 2--3 in the range $A=50$--$73$.
Therefore, the ratios $ \langle\rm ^{57}Ni/^{56}Ni\rangle $ and $ \langle\rm
^{58}Ni/^{56}Ni\rangle $ 
are suject to a considerable change in our models. 

\begin{figure}
\begin{center}
   \leavevmode\psfig{figure=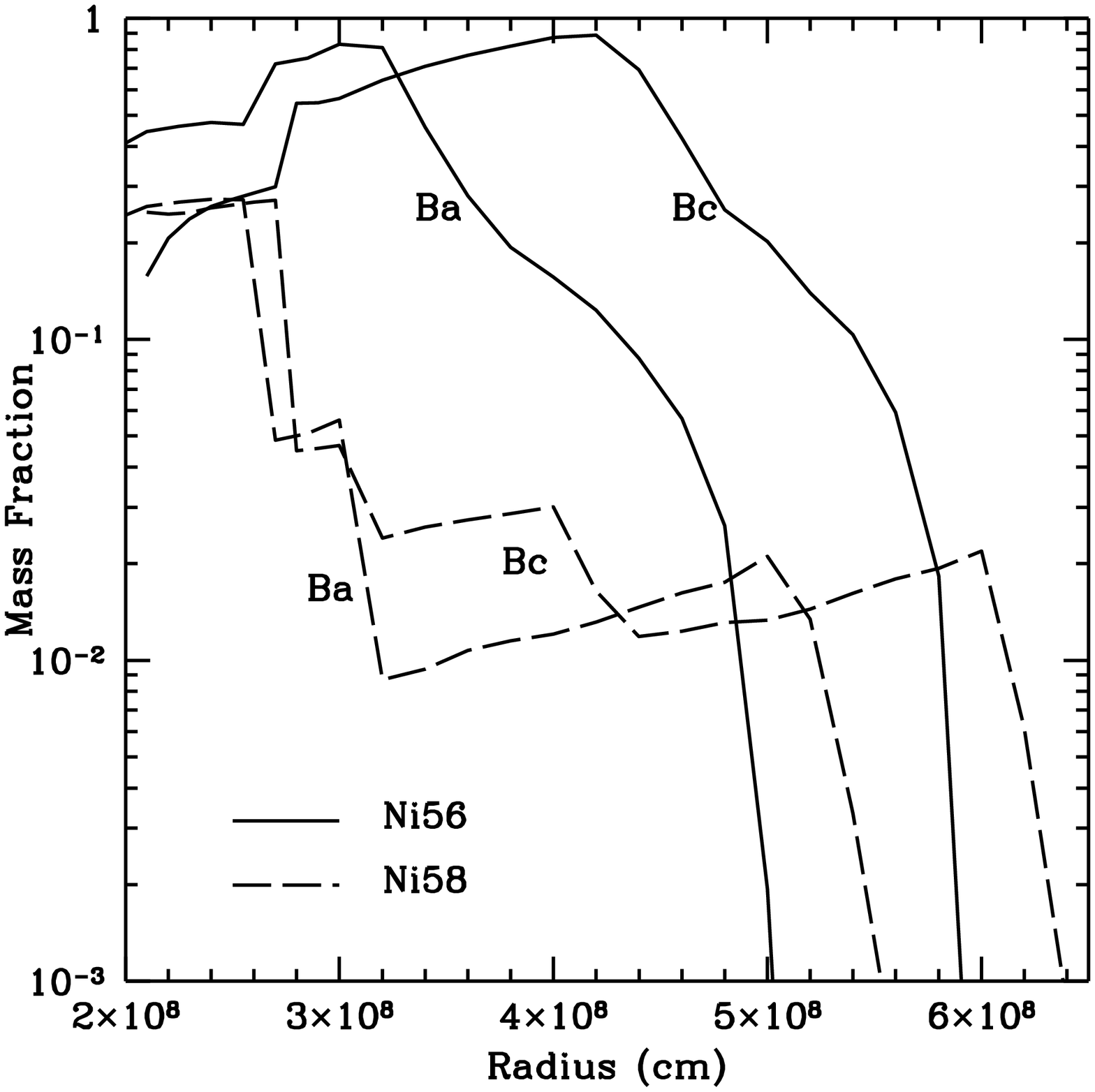,height=18cm,angle=0}
\end{center}
\caption{}
\label{massfrac}
\end{figure}

We show in table 4 the ratios $ \langle\rm ^{57}Ni/^{56}Ni\rangle $
and $ \langle\rm ^{58}Ni/^{56}Ni\rangle $ for all 
the 
models. 
We can see the tendency that the 
ratios 
$ \langle\rm ^{57}Ni/^{56}Ni\rangle $ and
$ \langle\rm ^{58}Ni/^{56}Ni\rangle $ 
become smaller 
as the ratio of the initial kinetic
energy to the internal energy 
becomes larger. 
The reason can be
explained 
as follows. 
The peak temperature reaches becomes in a model which has more kinetic
energy initially, and $\rm ^{56} Ni$ is more produced in the O-rich
layer. Therefore, the mass cut must be taken
larger in its size compared with the bomb method, and
neutron-rich elements, such as $\rm ^{57} Ni$ and $\rm
^{58} Ni$, are less included in the ejecta. 
To clarify this explanation, we show in figure~\ref{massfrac} the mass
fraction of $\rm ^{56}Ni$ and $\rm ^{58}Ni$
for 
models of Ba and Bc. It is evident that $\rm ^{56}Ni$ is
synthesized more in the outer region in 
model Bc 
compared with 
model Ba.
Additionally, we can see the tendency that the abundance ratios 
are smaller, since
the energy-deposition region is assumed to be smaller and to be located
farther away from the center. This is so for the same reason as
mentioned above.
As a consequence, the ratios of $ \langle\rm ^{57}Ni/^{56}Ni\rangle $
and $ \langle\rm ^{58}Ni/^{56}Ni\rangle $ are near
to the range of the observational uncertainty for model Cc, even if
the $Y_{\rm e}$ distribution of the progenitor is not changed artificially. 

We
comment on the total mass of $\rm
^{56}Ni$ synthesised in the O-rich layer. In 
this 
layer,
electron capture hardly occurs, contrary to the case in
the Si-rich layer.
If $\rm ^{56}Ni$ is synthesised mainly in the O-rich layer, the
treatments of electron capture and convection 
are
less important
concerning the outcome  
of explosive nucleosynthesis. 
As can be seen in table 4,
most of the $\rm ^{56}Ni$ is synthesized in the O-rich layer 
in model Cc. 

%We must note that these values are not good 
%to explain the bolometric light curve around $\sim
%1000$ days (Kumagai $et \; al$. 1993).
%%%%%%%%%%%%%%%%%%%%%%%%%%%%%%%%%%%%%%%%%%

%%%%%%%%%%%%%%%%%%%%%%%%%%%%%%%%%%%%%%%
\section{Summary and Discussion} \label{summary}
%%%%%%%%%%%%%%%%%%%%%%%%%%%%%%%%%%%%%%%

\indent

We have carried out 1-dimensional calculations using various initial
conditions in order to determine its influence on explosive nucleosynthesis.
We find a tendency that the peak temperature becomes higher if
the input energy is deposited 
more in the form of
kinetic energy than that of the internal energy, because this 
makes the mass cut larger; neutron-rich
matter is less included in the ejecta, which is good for reproducing
 the observational data. We must only say that a fairly 
strong initial shock-wave, like model Cc, must be assumed for reproducing 
the observation $ \langle\rm ^{58}Ni/^{56}Ni\rangle $, to be sure.

Since there is no reason why the input energy is deposited 
in the form of either internal energy or kinetic energy alone,  
it will be very favorable to find that 
some combinations of them can reproduce the observational data 
with the electron fraction of the progenitor unchanged. 
In other words, our conclusion is that treatments of the various
processes for stellar evolution, such as
convection and electron capture during 
the silicon burning stage, are still compatible with the results
of explosive nucleosynthesis
calculations. Then, the condition for
the initial shock wave in our models sould be the outcome 
of the core-collapse, the bounce, and shock-wave propagation
from the central region.
In particular, the demand concerning the amount of kinetic energy at
the Si-rich layer may be a strong constraint for the model of a delayed
explosion, because in that model the shock wave stalls at first and is
revived by neutrino heating, which may result in a low ratio of
the kinetic energy to the thermal energy.

We note the astrophysical implications  
of our results.
First, the central compact object could be a pulsar 
for all our models, since
the mass cut is below the upper limit of the neutron- 
star mass, even in the 
case of Cc. However, there is also the possibility to generate 
a more massive accretion disk around a neutron star 
for a larger mass cut. 
This 
means 
that the central
compact system has more activity in the 
larger 
mass cut case.
Secondly, more radioactive nuclei will be included in the compact
object if they 
survive 
there. This also shows a higher activity at
the central compact system, as suggested by Mineshige et al.
(1992). In particular, these effects 
lead to the emission 
of gamma-rays, which could contribute to the  observation of the
supernova remnant. 
To say conversely, 
the observation of gamma-rays from a pulsar
and its accretion disk may be a touchstone of our results.
Finally, we note the uncertainty of the abundance of nuclei in the
range $A=50$--$73$. Since the solar-system abundances of the heavy elements $A 
\ge 50$ are generated mainly by Type Ia supernova (Tsujimoto et al.
1995), this uncertainty
may be less stressed. However, we believe that we must keep this uncertainty
in mind when an indivisual collapse-driven supernova is observed and analyzed.

We comment on the hydrodynamical code.
Though our code is Eulerian and that of Aufderheide et
al. (1991) is Lagrangian, our results are consistent with theirs.
Therefore, our conclusions that the peak temperature
becomes higher, the mass cut 
becomes
larger, and neutron-rich matter
is less included in the ejecta do not depend on the hydrodynamical 
scheme.
Therefore, we believe that the uncertainty of the initial condition discussed
in the present research
should be kept in mind when studying explosive nucleosynthesis, and
that it would help to solve the problem concerning the overproduction
of neutron-rich
matter for some progenitor models.

\vskip1.0cm
This research has been
supported in part by a Grant-in-Aid for the Center-of-Excellence (COE) 
Reserch (07CE2002) and for the Scientific Reserch Fund (05243103,
07640386, 3730) of the Ministry of Education, Science, Sports and
Culture in Japan.
and by Japan Society for the Promotion of Science Postdoctoral
Fellowships for Research Abroad. 

\clearpage
\section*{References}
%\small
\re
Aitken D.K., Smith C.H., James S.D., Roche P.F., Hyland
A.R., McGregor P.J.\ 1989, MNRAS 235, 19
\re
Arnett W.D.\ 1996 Supernova and Nucleosynthesis (Princeton
University~Press, Princeton) chapter 9
\re
Aufderheide M.B., Baron E., Thielemann F.K.\ 1991, ApJ 370, 630
\re
Danziger I.J., Lucy L.B., Bouchet P.\ 1991 in Supernovae, ed
S.E. Woosley (Springer, Berlin)p69
\re
Fuller G.M., Fowler W.A., Newman M.\ 1980, ApJS 42, 447
\re
Fuller G.M., Fowler W.A., Newman M.\ 1982, ApJS 48, 279
\re
Hashimoto M.\ 1995, Prog. Theor. Phys. 94, 663
\re
Hashimoto M., Nomoto K., Shigeyama T.\ 1989, A\&A 210, L5
\re
Kurfess J.D., Johnson W.N., Kinzer R.L., Kroeger R.A., Strickman M.S., 
Grove J.E.\ 1992, ApJ 399, L137
\re
Meikle W.P.S., Allen D.A., Spyromilio J., Varani G.-F.\ 1989, MNRAS
238, 193
\re
Mineshige S., Nomoto K., Shigeyama T.\ 1992, A\&A 267, 95
\re
Nagataki S., Hashimoto M., Sato K., Yamada S.\ 1997, ApJ 486, 1026
\re
Nomoto K., Hashimoto M.\ 1988, Phys. Rep. 163, 13
\re
Rank D.M., Pinto P.A., Woosley S.E., Bregman J.D.,
Witterborn F.C., Axelrod T.S., Cohen M.\ 1988, Nature 331, 505
\re
Roe P.L.\ 1981 J. Comput. Phys. 43, 357
\re
Sedov L.I.\ 1959, Similarity and Dimensional Methods in
Mechanics (Academic~Press, New York) chapter 4
\re 
Shigeyama T., Nomoto K., Hashimoto M.\ 1988, A\&A 196, 141
\re
Thielemann F.-K., Hashimoto M., Nomoto K.\ 1990, ApJ 349, 222
\re
Tsujimoto T., Nomoto K., Hashimoto M., Thielemann F.-K.\
1994, in Evolution of the Universe and its Observational Quest,
ed ~K. Sato (Universal Academy Press, Tokyo) p553
\re
Witterborn F.C., Bregman J.D., Wooden D.H., Pinto P.A.,
Rank D.M., Woosley S.E., Cohen M.\ 1989, ApJ 338, L9
\re
Woosley S.E.\ 1988, ApJ 330, 218
\re 
Woosley S.E., Weaver T.A.\ 1988, Phys. Rep. 163, 79
\re
Woosley S.E., Weaver T.A.\ 1995, ApJS 101, 181
\re
Yamada S., Sato K.\ 1994, ApJ 434, 268

\bigskip
\bigskip
\bigskip

\newpage

\begin{table*}
\begin{center}
Table~1. \hspace{4pt} Radius of the interface for each layer.\\
\begin{tabular}{c|ccccccccccc}
\hline
\hline
Interface  &  Radius [cm] & Radius [$M_{\odot}$] \\  
\hline
Fe/Si & $1.5 \times 10^8 $ & $1.4$ \\
Si/O  & $3.0 \times 10^8 $ & $1.7$ \\
O/He  & $6.3 \times 10^9 $ & $3.8$ \\
\hline
\hline
\end{tabular}
\end{center}
\end{table*}

\begin{table*}
\begin{center}
Table~2. \hspace{4pt} Initial conditions and name of each model.\\
\vspace{6pt}
\begin{tabular*}
{12cm}{@{\hspace{\tabcolsep}
\extracolsep{\fill}}p{6pc}ccccccccccccccccc}
\hline
\hline
$\rm E_{kin} \; : \; E_{th}$$^{*}$  & & $\rm R_{inp} / 10^8 cm$$^{\dagger} $ &  \\
\hline
  &  1.0--1.5 & 1.4--1.5 & 1.9--2.0 \\  
\hline
0$\%$ : 100$\%$    &  Aa  & Ba  &  Ca  \\
30$\%$ : 70$\%$    &  Ab  & Bb  &  Cb  \\
70$\%$ : 30$\%$    &  Ac  & Bc  &  Cc  \\
\hline
\hline
\end{tabular*}
\vspace{6pt}\par\noindent
$*$ The
ratio of the initial kinetic energy to the internal energy.\\
$\dagger$ The radius ($10^8$ cm) where the input energy is deposited.\\
\end{center}

\end{table*}

\begin{table*}
\begin{center}
Table~3. \hspace{4pt} Mass cut and peak temperature at Si/O interface
for each model. \\
\begin{tabular}{cccccccccccc}
\hline
\hline
Model  & Mass cut ($M_{\odot}$) & Mass cut (cm) & Temperature (K) at Si/O \\  
\hline
Aa    & 1.54 & 1.97 $\times 10^8$ & 4.8 $\times 10^9$ \\
Ab    & 1.57 & 2.13 $\times 10^8$ & 5.2 $\times 10^9$ \\
Ac    & 1.58 & 2.21 $\times 10^8$ & 5.4 $\times 10^9$ \\
Ba    & 1.56 & 2.07 $\times 10^8$ & 5.3 $\times 10^9$ \\
Bb    & 1.59 & 2.28 $\times 10^8$ & 5.5 $\times 10^9$ \\
Bc    & 1.60 & 2.34 $\times 10^8$ & 5.8 $\times 10^9$ \\
Ca    & 1.60 & 2.33 $\times 10^8$ & 5.9 $\times 10^9$ \\
Cb    & 1.61 & 2.43 $\times 10^8$ & 6.3 $\times 10^9$ \\
Cc    & 1.63 & 2.56 $\times 10^8$ & 6.6 $\times 10^9$ \\
\hline
\hline
\end{tabular}
\end{center}
\end{table*}

\begin{table*}
\begin{center}
Table~4. \hspace{4pt} The ratios of $ \langle\rm ^{57}Ni /  ^{56}Ni
\rangle $ and $ \langle\rm ^{58}Ni /  ^{56}Ni \rangle $ for each model.
\begin{tabular*}
{18cm}{@{\hspace{\tabcolsep}
\extracolsep{\fill}}p{6pc}ccccccccccccccccc}
\hline
\hline
Model  & $ \langle\rm ^{57}Ni /  ^{56}Ni \rangle $ ($1.5 \pm 0.5$) & $ \langle\rm
^{58}Ni /  ^{56}Ni\rangle $ (0.7--1.0) 
& Mass of $\rm ^{56}Ni$ in O-layer \\
\hline
Aa    &  2.7  & 9.0  &  5.6 $\times 10^{-3} M_{\odot}$\\
Ab    &  2.3  & 6.6  &  2.3 $\times 10^{-2} M_{\odot}$\\
Ac    &  2.1  & 5.7  &  2.9 $\times 10^{-2} M_{\odot}$\\
Ba    &  2.3  & 6.2  &  1.7 $\times 10^{-2} M_{\odot}$\\
Bb    &  1.9  & 3.8  &  3.3 $\times 10^{-2} M_{\odot}$\\
Bc    &  1.8  & 3.0  &  3.8 $\times 10^{-2} M_{\odot}$\\
Ca    &  2.1  & 4.6  &  4.0 $\times 10^{-2} M_{\odot}$\\
Cb    &  2.0  & 3.5  &  4.7 $\times 10^{-2} M_{\odot}$\\
Cc    &  1.6  & 1.3  &  5.6 $\times 10^{-2} M_{\odot}$\\
\hline
\hline
\end{tabular*}
\end{center}
\vspace{6pt}\par\noindent
The values in the parentheses mean the
observational values. The last column denotes the total mass of $\rm
^{56}Ni$ synthesized in O-rich layer.
\end{table*}

\clearpage
\centerline{Figure Captions}
\bigskip
\begin{fv}{1}
{7cm}
{
Table of nuclei included in our nuclear reaction network;
242 species are included. The gray-colored nuclei denote stable nuclei.
}
\end{fv}
\begin{fv}{2}
{7cm}
{
Time variation of ($\rho,T$) for the test particle,
which is located at $r=3.0 \times 10^8$ cm. The dashed line is for
model Ba and the solid line is for model Bc.
}
\end{fv}
\begin{fv}{3}
{7cm}
{
Peak temperature of models Ba and Bc.
The abscissa means the initial position of each test particle.
The dashed line is for model Ba and the solid line is for model Bc.
}
\end{fv}

\begin{fv}{4}
{7cm}
{
Comparison of the composition for the mass number
range $A=16$--$73$ between models Ba and Bc.
}
\end{fv}

\begin{fv}{5}
{7cm}
{
Comparison of the abundance of each nucleus with the
solar value (normalised to $\rm ^{16}O$).
The open circles and solid points correspond to Ba/solar and
Bc/solar, respectively. 
}
\end{fv}

\begin{fv}{6}
{7cm}
{
Final mass fraction of $\rm ^{56}Ni$ and $\rm ^{58}Ni$
for models Ba and Bc. The radius means the initial radius of the progenitor.
}
\end{fv}

%\vspace*{10cm}

\newpage

\thispagestyle{empty}
\newpage

\thispagestyle{empty}
\newpage

\thispagestyle{empty}
\newpage

%\thispagestyle{empty}
%\begin{figure}
%\begin{center}
%   \leavevmode\psfig{figure=maxvel.ps,height=18cm,angle=0 }
%\end{center}
%\caption{}
%\label{maxvel}
%\end{figure}
%\newpage

\thispagestyle{empty}
\newpage

\thispagestyle{empty}
\newpage

\thispagestyle{empty}
\newpage

\end{document}